# Graphene-enabled, directed nanomaterial placement from solution for large-scale device integration

Michael Engel[1], Damon B. Farmer[2], Jaione Tirapu Azpiroz[1], Jung-Woo T. Seo[3], Joohoon Kang[3], Phaedon Avouris[2], Mark C. Hersam[3], Ralph Krupke[4,5,6], Mathias Steiner[1,2*]

[1] *IBM Research, Rio de Janeiro, RJ 22290-240, Brazil*

[2] *IBM Research, Yorktown Heights, NY 10598, USA*

[3] *Department of Materials Science and Engineering and Department of Chemistry, Northwestern University, Evanston, IL 60208, USA*

[4] *Institute of Nanotechnology, Karlsruhe Institute of Technology, 76021 Karlsruhe, Germany*

[5] *DFG Center for Functional Nanostructures (CFN), 76028 Karlsruhe, Germany*

[6] *Institut für Materialwissenschaft, Technische Universität Darmstadt, 64287 Darmstadt, Germany*

[*]*msteine@us.ibm.com, mathiast@br.ibm.com*

**Controlled placement of nanomaterials at predefined locations with nanoscale precision remains among the most challenging problems that inhibit their large-scale integration in the field of semiconductor process technology. Methods based on surface functionalization**[1] **have a drawback where undesired chemical modifications can occur and deteriorate the**

**deposited material. The application of electric-field assisted placement techniques[2] eliminates the element of chemical treatment; however, it requires an incorporation of conductive placement electrodes that limit the performance, scaling, and density of integrated electronic devices[3]. Here, we report a method for electric-field assisted placement of solution-processed nanomaterials by using large-scale graphene layers featuring nanoscale deposition sites. The structured graphene layers are prepared via either transfer or synthesis on standard substrates, then are removed without residue once nanomaterial deposition is completed, yielding material assemblies with nanoscale resolution that cover surface areas larger than 1mm$^2$. In order to demonstrate the broad applicability, we have assembled representative zero-, one-, and two-dimensional semiconductors at predefined substrate locations and integrated them into nanoelectronic devices. This graphene-based placement technique affords nanoscale resolution at wafer scale, and could enable mass manufacturing of nanoelectronics and optoelectronics involving a wide range of nanomaterials prepared via solution-based approaches.**

Bottom-up, large-scale manufacturing of integrated electronics requires application of substrate patterning, chemical surface functionalization, and/or Langmuir-Blodgett-type techniques for depositing solution-based, semiconducting materials at the substrate's surface[4]. Chemistry-based techniques can be applied to deposit nanomaterials at wafer scale[5], however, they offer limited control over the material's location, orientation, and density. In addition, the use of aggressive chemicals may lead to inferior device performance once the integration process is completed.

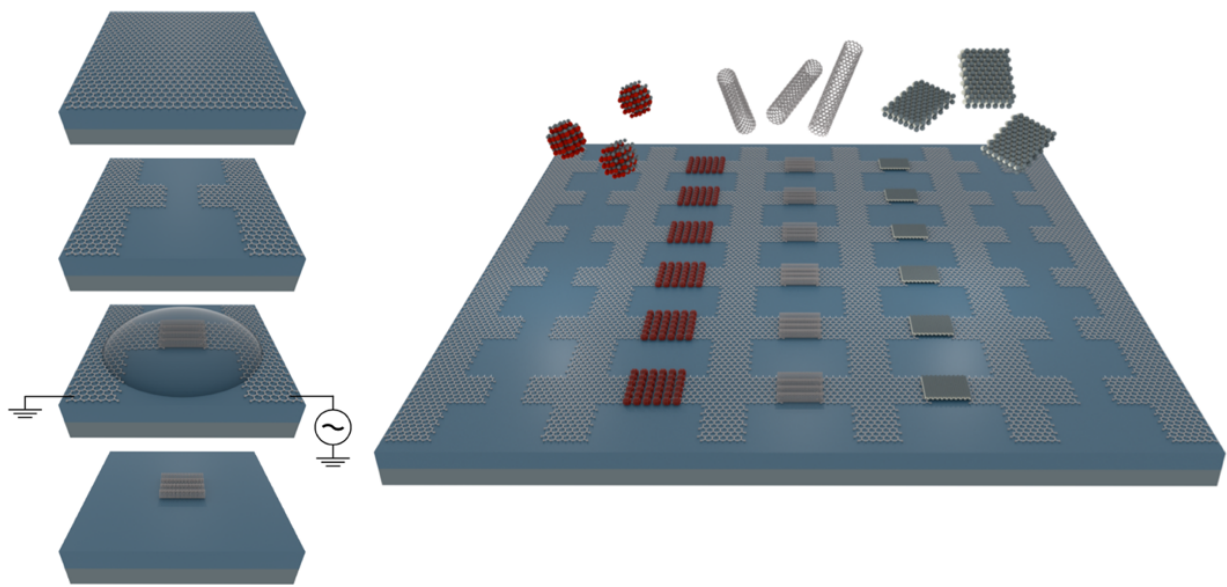

**Figure 1. Electric-field assisted placement of nanomaterial from solutions with patterned graphene.** Artistic rendering of electric field assisted placement of nanoscale materials between pairs of opposing graphene electrodes structured into a large graphene layer located on top of a solid substrate. The sketches on the left hand side visualize the key steps of the method: structuring a large-scale graphene layer to form local electrodes, applying a AC voltage for facilitating the field-assisted placement of nanomaterials from solution (shown are carbon nanotubes), and removing the graphene structure once the placement is complete. In the visualization on the right hand side, quantum dots (0D), single-walled carbon nanotubes (1D), and layers of molybdenum disulfide (2D) are shown as representative nanomaterials that can be assembled at large scale based on the graphene-based, electric field assisted placement method.

As an alternative method, the application of electric field assisted deposition requires the presence of conductive electrodes for generating an attractive force that drags nanomaterials to predefined locations where material deposition is desired[6]. The method was applied successfully to integrate low-dimensional semiconductors in exploratory electronic and optoelectronic devices; e.g. zero-dimensional quantum dots (0D)[7], one-dimensional nanowires and nanotubes (1D)[8,9], and two-dimensional graphene[10,11]. However, due to the inability to remove the conductive electrodes after deposition, they remain at the device locations, limiting electronic device performance and integration density[3]. In order to harvest the nanoscale placement potential of the dielectrophoresis method at large scale and, at the same time, to avoid the use of unwanted conductive electrodes in the placement process, another solution is needed. Graphene, an excellent candidate material for electric-field assisted material deposition, offers a potential solution to this, as it can be removed residue-free in a standard etching procedure[12]. In addition, this material possesses a combination of properties critical to this placement technique, including AC operation capability[13], wafer scale growth[14,15] or transfer onto a wide range of standard substrates[16], and nanometer resolution patterning[17].

In Figure 1, the principle of the electric-field assisted placement method using graphene electrodes is visualized for three representative nanomaterials. A large graphene layer is patterned to form sets of opposing electrodes at predefined positions on top of a solid substrate. The application of an alternating voltage between the graphene layers generates local electric-fields at predefined substrate positions. If nanomaterial dispersions are placed on top of the substrate, a dielectric force attracts the nanomaterials towards the positions defined by the graphene electrodes where the nanomaterials settle at the substrate surface. Finally, the graphene layers are removed leaving behind the nanomaterials assembled at the predefined substrate positions. In addition to its excellent properties as an AC conductor[18], graphene offers optimal geometrical placement conditions for nanomaterials in comparison with the much thicker metallic electrodes typically used in electric-field assisted deposition methods[8].

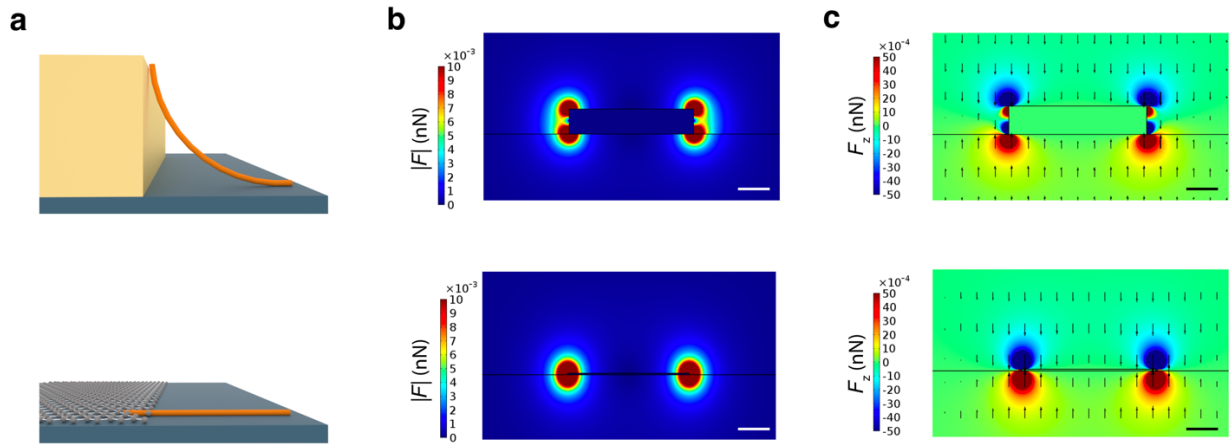

**Figure 2. Geometrical and electrostatic conditions in graphene-based placement of nanomaterials. a** Cross section of a standard metal electrode used for electric field assisted assembly of nanomaterials and a graphene layer used for the same purpose, respectively, both located on top of a solid substrate. As a reference placement example, a carbon nanotube is shown that bridges contact and substrate. Cross section of the **b** dielectrophoresis force distribution and **c** dielectrophoresis force z-component exerted on a carbon nanotube in aqueous solution which is generated by a dc voltage applied to the standard metal electrode (top) and the graphene layer (bottom), respectively. Scale bar: 50nm.

In Figure 2, we visualize the geometrical cross-sections and the electric field distributions at the electrode's contact edge. As an ideal 2D nanomaterial with atomic thickness, graphene enables reduction of the thickness of conventional deposition electrodes by two orders of magnitude, from several tens of nanometers (metal electrodes) to less than one nanometer. The electric field distribution is strongly localized at the substrate surface, providing ideal deposition conditions in close vicinity of the graphene layer edge. In comparison with a standard metal electrode, the deposited nanomaterial exhibits significantly lower bending at the contact edge, which reduces the likelihood of device breakdown[19,20].

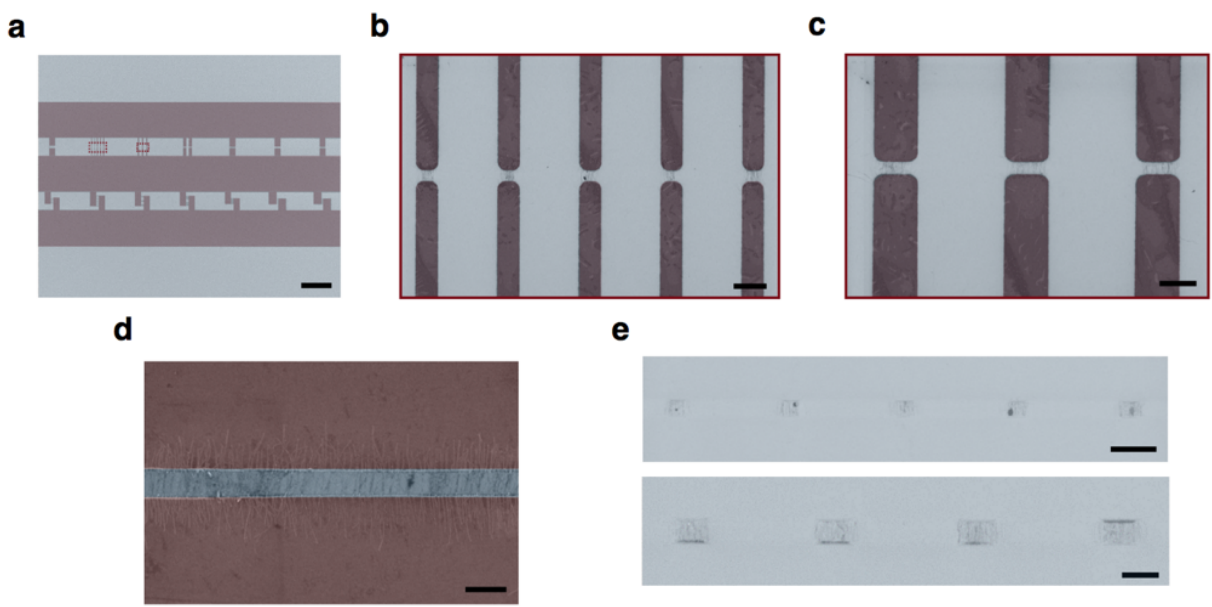

**Figure 3. Pitch and orientation in graphene-based, electric field assisted nanomaterial placement a**. Scanning electron microscope (SEM) false color image of a graphene layer on a SiC substrate featuring test structures of varying size and orientation for AC field-assisted assembly of carbon nanotubes from solution. The dashed lines frame the areas that are shown in b, c. Scale bar: 50μm. **b, c** Magnified SEM false color images taken at the positions highlighted by dashed frames in a showing representative graphene deposition structures after carbon nanotube placement is completed**.** Scale bars: 2μm. **d** SEM false color image of an extended graphene electrode pair after placement is completed, exhibiting position and orientation of carbon nanotubes with respect to the gap. Scale bar: 1μm. **e** SEM false color images of carbon nanotube assemblies after removal of the graphene layer. The upper panel corresponds to the area imaged in b. Scale bar: 2μm.

An example of a structured graphene layer for electric field assisted nanomaterial deposition is shown in Figure 3. A large-scale graphene layer (mm scale) features smaller structures (μm to nm scale) designed for investigating deposition quality as a function of pitch. The microscopy images demonstrate the design flexibility for the graphene-based deposition electrodes, and details regarding their fabrication are provided in the Methods Section. In our experiments, we have covered the entire graphene layer system uniformly with aqueous solution containing spatially isolated, semiconducting carbon nanotubes as a reference material for nanomaterial deposition[21]. Upon application of the deposition voltage $V_{DEP}$, as visualized in Figure 1, carbon nanotube placement occurs simultaneously across the solution-covered substrate surface at locations framed by opposing graphene electrode pairs. Generally, the material density can be adjusted by tuning $V_{DEP}$ and the nanoparticle concentration in solution. For a given set of parameters, we obtain a constant and uniform placement density, largely independent of the size and distance between graphene electrodes. In the examples shown in Figure 3, a rather constant nanotube density of about 15μm$^{-1}$ is maintained. Importantly, no carbon nanotubes are found at positions outside the predesigned electrode areas. We therefore conclude that size and position of deposition sites are mainly determined by the graphene pattern. For analyzing the density and orientation of nanotubes within the deposition area, we have designed laterally extended electrode pairs like the one shown in Figure 3d. As expected from the electric field distribution and the nanotube geometry (see Figure 2) the carbon nanotubes align symmetrically across the gap with a density of about 15μm$^{-1}$ at an average angle of (90±10) degrees. We note that increasing either deposition voltage or concentration of nanoparticles in solution increases the material density in the placement, to values above 50μm$^{-1}$. Overall, this placement of individual nanotubes and nanotube bundles in aligned arrays with a homogeneous density is highly suitable for scalable manufacturing of high-performance electronic devices.

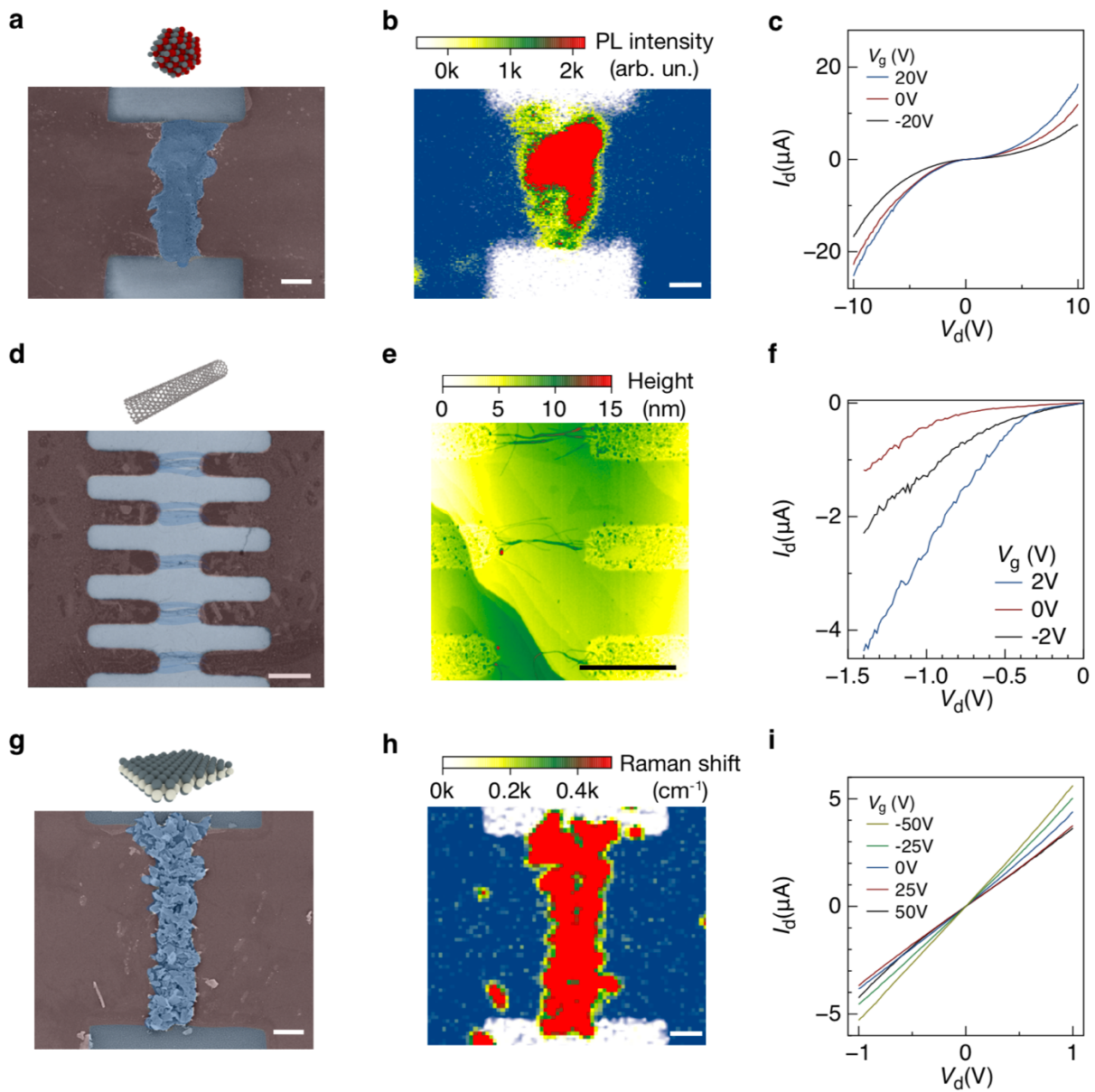

**Figure 4. Spatially resolved deposition, characterization and device integration of 0D, 1D, 2D semiconductors.**
**a** Scanning electron microscope (SEM) false color image of a CdSeS/ZnS quantum dot assembly. Scale bar: 1μm **b** Raman 2D intensity false color image (white: 20 arb. units, blue: 100 arb. units) indicating the graphene areas, overlaid by a photoluminescence intensity false color image spectrally integrated at (580±10)nm indicating the position of quantum dots. The images are taken at the same area as in a. **c** Electrical transport characteristic of a quantum dot thin film device. Scale bar: 1μm **d** SEM false color image of carbon nanotube assembly. Scale bar: 1μm. **e** Atomic force microscope image of the carbon nanotube assembly imaged in d. Scale bar: 1μm **f** Electrical transport characteristic of a carbon nanotube thin film device. **g** SEM false color image of few-layer molybdenum disulfide assembly. Scale bar: 1μm **h** Raman 2D intensity false color image (white: 20 arb. units, blue: 100 arb. units) indicating graphene areas, overlaid by a Raman intensity false color image spectrally integrated at $(385\pm10)\text{cm}^{-1}$ indicating position of few-layer molybdenum disulfide. The images are taken at the same area as in g. Scale bar: 1μm **i** Electrical transport characteristic of a molybdenum disulfide thin film device.

In order to investigate the broader applicability of the method, we show in Figure 4 the material placements for three representative solution-processed, low-dimensional (0D, 1D, and 2D) semiconductors, namely CdSeS/ZnS alloyed quantum dots, semiconducting carbon nanotubes, and few-layer $MoS_2$, respectively. The material is identified by *in situ* spectroscopic analysis and characterized by atomic force microscopy to confirm the chemical identity and surface coverage. The results demonstrate that the method allows effective deposition of solution-based semiconducting nanomaterials, regardless of their specific material properties. In order to determine the electrical transport properties of the nanomaterial deposits, we have manufactured three terminal field-effect devices; see Methods Section. By applying appropriate bias voltages to the devices, we obtain the expected semiconducting transport characteristics for all reference materials used in this study. The results suggest that the deposition method can, in principle, be used to manufacture electronic devices from a broad range of solution-based nanomaterials.

For investigating the integration and scaling potential of the placement method, we have applied the graphene-based material deposition to an integrated $SiO_2$/Si gate stack customized for three-terminal field-effect transistor integration (see Figure 5). The gate stack contains a regular array of embedded, metallic electrodes that enables downscaling of transistor dimensions while maintaining gate coupling and planarity. In the present example, the minimum gate electrode length is 40 nm. For the mass production of transistors in parallel, we have patterned the graphene layers to match the position of the embedded electrodes in the substrate (see Figure 5). In this manner, we are able to place the nanomaterials with a spatial resolution below 100 nm over areas with lateral lengths larger than 5 mm. Electrical transfer and output characteristics, of a transistor examined in a random sample from this series exhibit current on/off ratios of $10^5$ attesting to the absence of metallic carbon nanotubes in the device. While further research is needed to quantify the device yields by means of automated electrical transport analysis, we note that based on a visual inspection the deposition yields achieved in the current architecture are close to unity.

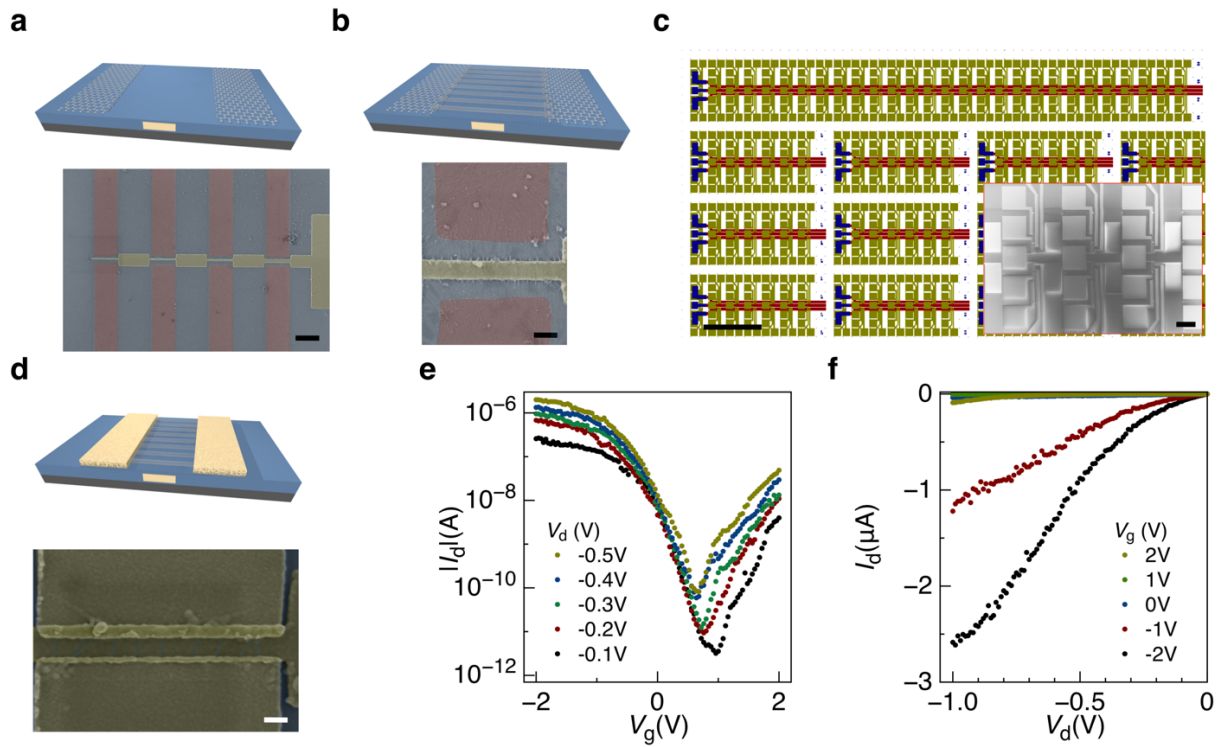

**Figure 5. Large-scale device integration and scaling with graphene-based, electric field assisted nanomaterial placement. a**. Artistic rendering of a Si/SiO$_2$ substrate with embedded metal electrode and a patterned graphene layer for field-assisted carbon nanotube placement. The scanning electron microscope (SEM) false color image reveals how the graphene layer is patterned to form gaps at the position of the embedded metal electrode for facilitating the nanotube placement. Scale bar: 1μm. **b**. Artistic rendering and SEM false color image of the carbon nanotube

placement across the embedded metal electrode, at the substrate surface, before the graphene electrodes are removed. Scale bar: 200nm. **c** Layout of the large scale integration based on the device architecture in a,b,d. The inset shows a SEM image of the dice-level implementation (top view). Scale bar: 50 μm. **d** Artistic rendering and SEM false color image (top view) showing final carbon nanotube array transistor, after removal of graphene layer and manufacturing of metal contacts. Scale bar: 100nm. **e** Electrical transfer characteristics of a carbon nanotube array transistor. **f** Electrical output characteristics of a carbon nanotube array transistor.

We note that large scale integration of high-performance electronic circuits made from nanotube solutions, as recently demonstrated in[22], is feasible based on the method reported here. While experiments in this work are performed at the scale of wafer dies, implementation at full wafer scale could soon be demonstrated, considering the feasibility of wafer scale dielectrophoretic assembly with standard metal electrodes[23]. In addition, our approach enables the design of integrated circuits such as integrated photodetector or light-emitting diodes circuit where certain device functionalities are carried out by different nanomaterials. In that regard we note that successive deposition steps are feasible with the method discussed here, i.e., once the materials are deposited, they remain adhered to the substrate surface and tolerate multiple subsequent deposition and rinsing steps. Another important direction of future research is the demonstration of the graphene-based placement method on a single particle level. Since dielectrophoresis with metal electrodes has already achieved individual nanotube and nanowire nanowires precision[8,9], an extension of our method could enable large-scale integration of exploratory quantum electronics and optoelectronics devices, e.g. [24].

In summary, we have reported a method for electric-field assisted placement of solution-processed nanomaterials by using patterned graphene layers that can be removed without residue after the deposition of nanomaterials is completed. This method is compatible with conventional semiconductor processing and can be applied to a broad class of nanomaterials and substrates used in industrial-scale production. It is amenable to further device scaling and complex integration processes, and can be simply extended to wafer-scale device manufacturing. Ultimately, we present a generalizable method that opens a route to bottom-up integration of nanomaterials for industry-scale integrated circuits.

**METHODS**

**Device Fabrication**

We implemented and tested the graphene-based placement method on (1) SiC and on (2) highly insulating Si with an $SiO_2$ capping layer having a thickness of 300nm. For (2), see the implementation example in Fig. 5 where we manufactured the $SiO_2$/Si substrates with local bottom gate electrodes. To that end, we first patterned a PMMA layer by e-beam lithography, followed by $SiO_2$ etching, metal deposition, and lift-off in hot acetone. In the next step, we deposited 30nm of $Al_2O_3$ at 250°C by using atomic layer deposition (ALD). We then transferred CVD-grown graphene[25] by a PMMA assisted technique[16] onto the gate stack. Starting the transfer by first spin coating PMMA (A4, 950 PMMA, MicroChem Corp.) on pre-cut pieces of CVD graphene on a copper foil (Graphene Laboratories, Inc.), we then placed the graphene on copper foil into copper etchant (PC Copper Etchant-200, Transene Company, Inc.) for 20min. After copper was dissolved, we transferred the floating PMMA/graphene bilayer into two subsequent water baths (high purity, de-ionized water, with a resistivity of 18MΩ-cm) to remove residual chemicals. The cleaned PMMA/graphene was then scooped onto the target substrate, blown dry with clean nitrogen, and baked at 100°C for several hours to remove residual water. We then defined graphene contact areas by patterning a negative tone resist bilayer (PMMA/HSQ), followed by an oxygen plasma step to etch exposed graphene areas. Subsequently, we defined large metallic contact pads by e-beam lithography. This step was followed by metal thin film evaporation (5nm Ti/50 nmAu), and a lift-off step. The larger contact pads were located on the edges of the sample to function as interfaces for the external measurement system. We then applied an alternating voltage (0.1-10MHz, 1-10$V_{p-p}$) to the graphene contacts by means of a waveform generator (Stanford Research Systems, DS345) via co-planar microwave probes (GGB Industries) operated in a ground-source-ground configuration. Upon application of the voltage, we deposited about 20µl of highly diluted nanomaterial solution onto the sample surface for periods of 1-10mins. The deposition process was completed by rinsing the sample sequentially with water and isopropyl alcohol. Following nanomaterial deposition, we defined a mask on top of the as-deposited material by patterning a negative tone resist bilayer (PMMA/HSQ), followed by an oxygen plasma step to etch the exposed graphene areas. In a final step, we patterned metallic contacts to the deposited nanomaterial, followed by metal evaporation (5nm Ti/50nm Au) and a lift-off step, for enabling electrical characterization with the external measurement setup.

**Nanomaterial Dispersion Preparation**

The semiconducting carbon nanotube solution was prepared as previously reported via density gradient ultracentrifugation, with >99% semiconducting purity and refined average diameter of 1.6nm[26-28]. Molybdenum disulfide ($MoS_2$) dispersion preparation began with 60mg of $MoS_2$ powder (American Elements) and 15mL ethanol-water co-solvent (10mL ethanol and 5mL deionized water) being placed in a 50mL plastic conical tube with a sealed setup[29] to avoid solvent evaporation and exfoliated by ultrasonication (Fisher Scientific Model 500 Sonic Dismembrator) at ~30W for 1hr in an iced bath. The resulting MoS$_2$ dispersions were then centrifuged at 5,000 rpm for 30min to remove unexfoliated MoS$_2$ powder (Avanti J-26 XP, Beckman Coulter) and the supernatant was carefully decanted. CdSeS/ZnS alloyed quantum dots in aqueous solution (COOH functionalized, fluorescence wavelength maximum $\lambda_{em}$ 575nm, 6nm diameter, 1mg/mL in H$_2$O) were purchased from Sigma-Aldrich.

**Material and Device Characterization**

In order to identify deposited nanomaterials *in situ*, we have acquired hyperspectral images of as-fabricated samples by combining Raman and photoluminescence micro-spectroscopies with a confocal laser scanning microscope (alpha300 RAS, WITec GmbH). In addition, we have performed in the same measurement system topographical characterization of the samples by means of atomic force microscopy. Electrical transport measurements have been performed under N$_2$ flow by using a probe station (FWP6, LakeShore) equipped with tungsten probes having a 50μm tip radius mounted on ceramic blade probe bodies, and a semiconductor parameter analyzer (B1500A Semiconductor Device Analyzer, Agilent).

**Electric Field Simulation**

We perform quasi-electrostatic field simulation using COMSOL Multiphysics (COMSOL Inc.) to evaluate the dielectrophoretic force field. We set up the simulation domain identical to the experimental conditions (physical dimensions, frequencies, applied voltages) using literature values for all relevant physical quantities (electrical conductivity, relative permittivity). Modeling the polarized bound charges of a rod-shaped particle immersed in an electric field as an induced dipole, we calculate the time-averaged dielectrophoretic force acting upon a carbon nanotube from the simulated electric field by the following relation[30]

$$\vec{F}_{DEP} = \frac{\pi d^2 l}{8} \varepsilon_m Re[CM] \nabla |\vec{E}_{rms}|^2$$

In this expression, $d$ and $l$ denote the diameter and length of the tube, respectively, *CM* represents the Clausius–Mossotti factor given as

$$CM = \frac{\varepsilon_t - \varepsilon_m}{\varepsilon_m + (\varepsilon_t - \varepsilon_m)L}$$

and $\varepsilon_t$ and $\varepsilon_m$ are the complex permittivity of the carbon nanotube and surrounding medium, respectively. The complex permittivity is given by $\varepsilon = \varepsilon_0 \varepsilon_r - i\sigma/\omega$, where $\varepsilon_r$ denotes the relative permittivity of the material, $\varepsilon_0$ the free space permittivity (8.85 $10^{-12}$ F/m), σ (S/m) is the conductivity and ω (rad/s) the angular frequency. Our simulations assume a depolarization factor $L$ equal to $10^{-5}$, a tube diameter $d$ equal to 1nm and a length $l$ equal to 500nm[31]. Under those conditions, all metallic and most semiconductor carbon nanotubes experience positive DEP (Re($CM$) > 0), i.e. the nanomaterial is attracted to electric field intensity maxima at the edges of the electric contact. For simplicity, a Clausius–Mossotti factor of 1 was used in the simulations.

**ACKNOWLEDGEMENT**

The authors thank Christos Dimitrakopoulos (University of Massachusetts, Amherst, Massachusetts, USA) for supplying graphene on SiC, Bruce Ek and Jim Bucchignano (both IBM Research) for expert technical assistance, and Ulisses Mello (IBM Research) for support. J.K., J.-W.T.S., and M.C.H acknowledge support from the National Science Foundation (DMR-1505849) for the preparation of nanomaterial dispersions.